# Graphene functionalised by laser ablated $V_2O_5$ as highly sensitive $NH_3$ sensor

Margus Kodu*, Artjom Berholts, Tauno Kahro, Mati Kook, Peeter Ritslaid, Helina Seemen, Tea Avarmaa, Harry Alles and Raivo Jaaniso

Address:

Institute of Physics, University of Tartu, W. Ostwald Street 1, EE50411 Tartu, Estonia

Email: Margus Kodu* – margus.kodu@ut.ee

* Corresponding author

## Abstract

Graphene has been recognized as a promising gas sensing material. The response of graphene-based sensors can be radically improved by introducing defects in graphene using, e. g., metal or metal oxide nanoparticles. We have functionalised CVD grown, single layer graphene by applying pulsed laser deposition (PLD) of $V_2O_5$ which resulted in a thin $V_2O_5$ layer on graphene with average thickness of ~0.6 nm. According to Raman analysis, PLD process also induced defects in graphene. Compared to unmodified graphene, the obtained chemiresistive sensor showed considerable improvement of sensing ammonia at room temperature. In addition, also the response time, sensitivity and reversibility were essentially enhanced due to graphene functionalisation by laser deposited $V_2O_5$. This can be explained by increased surface density of gas adsorption sites introduced by high energy atoms in laser ablation plasma and formation of nanophase boundaries between deposited V$_2$O$_5$ and graphene.

## Introduction

Graphene as ultimately thin (semi)conducting material is a promising gas sensing system. Highly sensitive responses down to a single-molecule-resolution have been demonstrated with graphene-based devices under laboratory conditions [1,2,3]. However, in order to develop gas sensing applications working under real conditions, a lot of efforts has been dedicated to modification of graphene for improving its gas sensing characteristics. Especially, increasing the selectivity of graphene-based gas sensors is crucial for their future implementation. Recently, improvement of gas sensing characteristics has been demonstrated with resistive type gas sensors based on a single layer graphene modified by a deposited layer of precious metal [4] or metal oxide nanoparticles [5]. Also, introduction of suitable defects was shown to have positive effect on gas adsorption and sensor properties of graphene [6].

Transition metal oxides constitute an important class of catalysts and photosensitizers. Apart from the very first and last 3d elements, scandium and zinc, the rest of metals possess several oxidation states. The presence of several stable oxidation states serves as a basis of catalytic activity in redox reactions and is most noticeable for vanadium, chromium, and manganese. In particular, vanadium has the highest oxidation state in vanadium pentoxide ($V_2O_5$) - a good oxygen transfer catalyst that is (thermally) stable in air and vacuum [7-10]. Therefore, we considered vanadium oxide as a promising material for functionalising of graphene sensor, in order to increase its selectivity towards reducing pollutant gases, such as ammonia. Vanadium oxide based films and nanostructured layers have been previously synthesised for gas sensing applications by various methods [11], including pulsed laser deposition (PLD) [12].

PLD is a highly versatile method for relatively well controlled preparation of thin films, and many advanced composite materials have been produced for diverse applications [13]. The possibility to evaporate practically any solid material, tune the kinetic energy of particles between 0.1 to 1000 eV, as well as the amount of deposited material from only about $1/100^{th}$ of a monolayer per laser pulse are the advantages worth mentioning. The method of PLD has recently been applied to improve the nitrogen dioxide ($NO_2$) sensing properties of chemical vapour deposition (CVD) grown, single layer graphene in our previous work, using $ZrO_2$ and Ag for functionalisation [14].

In the present work, we demonstrate functionalisation of single layer CVD graphene with a few layers of laser deposited $V_2O_5$. The amount and chemical state of vanadium oxide on graphene was characterized by X-ray photoelectron spectroscopy and X-ray fluorescence. The impact of PLD process on graphene defect structure was investigated using Raman spectroscopy. Basing on the electric conductivity modulation, the room temperature gas sensing properties of the manufactured sensor structure towards ammonia ($NH_3$) and (for comparison) nitrogen dioxide ($NO_2$) gases were investigated.

# Results

Figure 1(a) shows a typical Raman spectrum from graphene placed between the electrodes of gas sensor structure (see Experimental section). The G and 2D bands peak at ~1590 cm$^{-1}$ and ~2690 cm$^{-1}$, and have full-widths at half-maximum of 11 cm$^{-1}$ and 29 cm$^{-1}$, respectively. These characteristics, together with the ratio of G to 2D band peak intensities of approximate 1:3 correspond to a single layer graphene [15]. The absence of defect-related D band at ~1350 cm$^{-1}$ indicates extremely low defect density [15, 16]. After the PLD of $V_2O_5$ onto graphene (for the PLD details, see the Experimental section below) the defect-related D and D´ bands emerge in Raman spectrum, and, at the same time, the G and 2D bands decrease in size (Figure 1(b)).

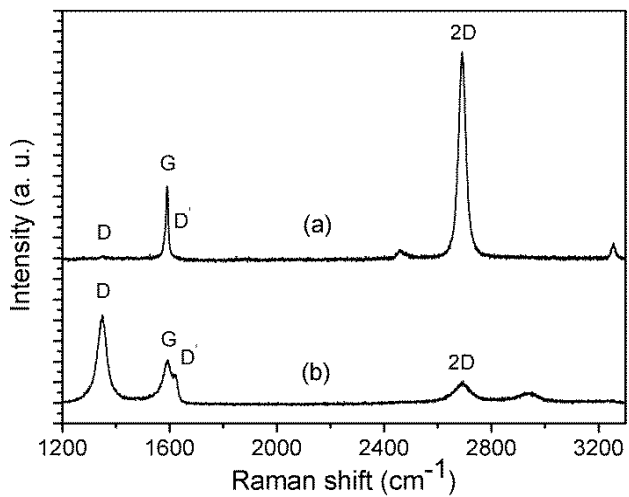

**Figure 1:** Typical Raman spectra of the graphene sensor device recorded between electrodes (a) before and (b) after laser deposition of $V_2O_5$. Defect related peaks D and D' are absent in (a).

Figure 2 shows scanning electron microscope (SEM) images of graphene surface (a) before and (b) after PLD of $V_2O_5$. The darker contrasting regions and the lines in the image originate from the Cu growth surface of CVD process or wrinkles left in the graphene sheet during the transfer process from the copper foil to Si/SiO$_2$ substrate. These features are characteristic for CVD graphene and can also be seen in the SEM image of pristine graphene shown in Figure 2a. The islands of about ~20 nm in diameter of laser deposited nanostructured material can be distinguished in the image. It is well known that gas phase species created by laser ablation of solids have a wide distribution of kinetic energy [17]. A considerable fraction of particles can have sufficient energy (~100 eV) for creation of point defects in the graphene sheet [18]. As a consequence, the Raman lines assignable to point defects or imperfect graphene edges appear. The extent of disorder induced by the PLD process is characterized by the ratio of Raman line intensities $I_D/I_G$. Considering the ratio $I_D/I_G$ = 2.13 obtained from Figure 1, and applying the formulae given by Piment et al. [19] and Concado, et al. [20], estimates are obtained for average graphene crystallite size and the distance between point defects of ~7.9 nm and ~7.7 nm, respectively.

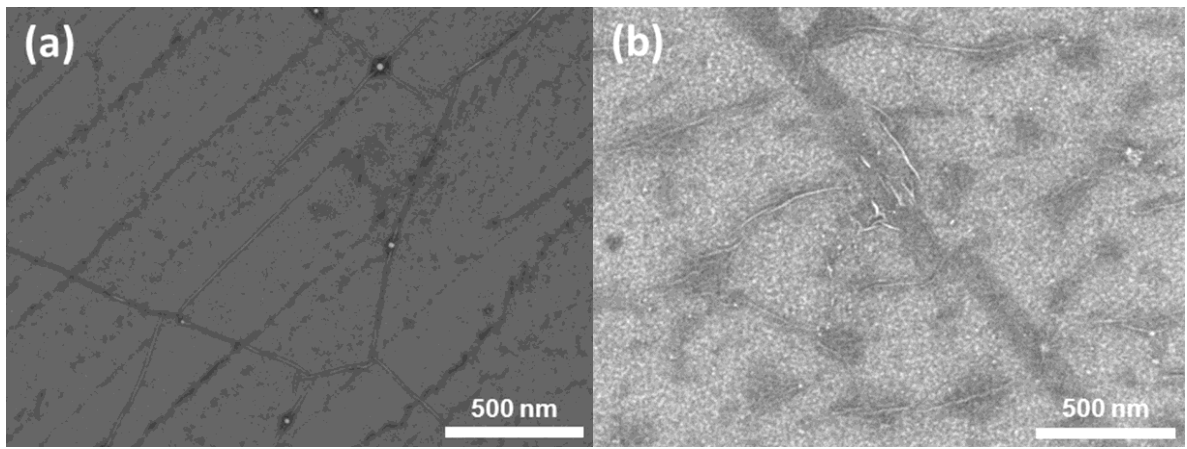

**Figure 2**: SEM images of pristine single layer graphene surface (a) and graphene functionalised by V$_2$O$_5$ (b).

The amount of vanadium that was deposited on the sensor substrate was evaluated using X-ray fluorescence (XRF) analysis, by measuring the amount of V in the film deposited onto fused quartz substrate in identical PLD procedure. According to XRF, the mass thickness of element V on the substrate was 0.11 µg/cm$^2$. Presuming that all of vanadium on the substrate belongs to composition $V_2O_5$, and considering the density of crystalline $V_2O_5$ of 3.38 g/cm$^3$ [21], an estimate for the average thickness of continuous $V_2O_5$ layer on graphene is 0.58 nm. In other words, the approximate number of $V_2O_5$ (001) lattice plane layers is about 2.5.

The oxidation state of vanadium was determined by means of X-ray photoelectron spectroscopy (XPS). Figure 3 depicts the oxygen 1s and V 2p regions of XPS spectrum measured from the surface of sample after the PLD process. Because the source was not monochromatic, the V 2p$_{1/2}$ region is not usable for the analysis, owing to the overlap with the Mg K$_\beta$ satellite with O 1s peak. The main component of V 2p$_{3/2}$ at binding energy 517.4 eV was assigned to $V^{5+}$, whereas a faint component at 513.3 eV belongs to $V^{2+/1+}$. The O 1s peak component at 530.9 eV was identified as oxygen bound to vanadium, in agreement with Biesinger's XPS measurements on vanadium oxide [22]. The main peak component of O 1s at 533.1 eV was attributed to oxygen of $SiO_2$ in silica substrate that can possibly overlap with the C–O component of (partly oxidized) graphene at the same binding energy (533.1 eV) [23]. Thus, according to XPS and XRF analysis, the material deposited on graphene is predominately $V_2O_5$, with an average thickness of about 0.6 nm. Among other vanadium oxides, $V_2O_5$ is a relatively stable compound that can be easily deposited by PLD method [12].

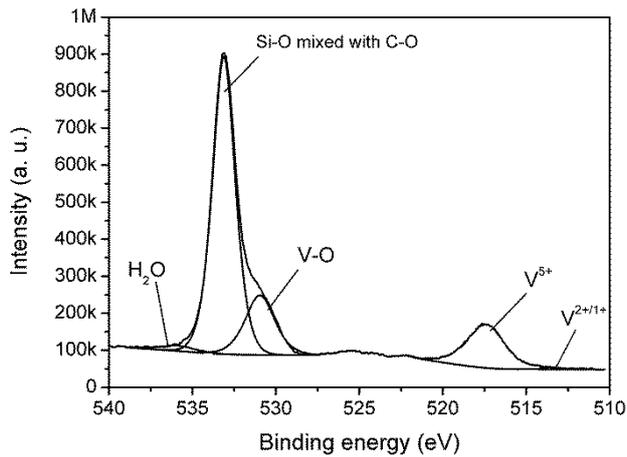

**Figure 3**: XPS spectra of graphene following PLD treatment with $V_2O_5$ in the O 1s and V 2p region. Non-monochromatic Mg Kα was used for excitation and the satellite peaks were removed.

Figure 4 depicts the time responses of PLD-functionalised graphene sensor to polluting gases $NO_2$ and $NH_3$. All the gas measurements in this work were recorded under continuous excitation with ultraviolet (UV) light (λ = 365 nm) at room temperature (RT). Illumination by UV light can enhance the sensing performance of graphene based gas sensors. Increased gas response and fast recovery under the UV irradiation has been demonstrated for pristine and functionalised graphene based sensors, possibly as a result of cleaning the surface from interfering or passivating gases [24-26]. The effect has been explained by photo-induced desorption of oxygen and water molecules, thus activating additional adsorption sites on graphene for the target gas.

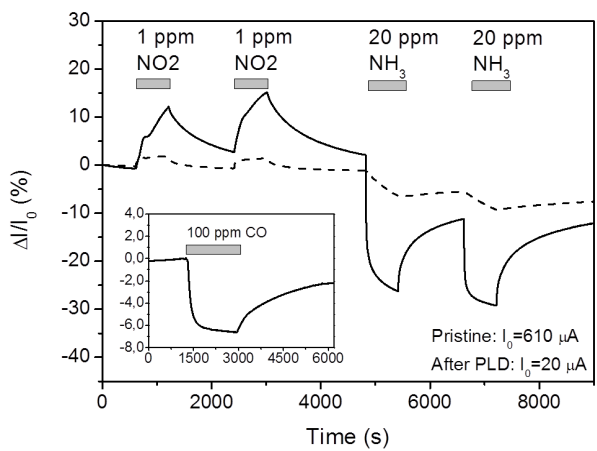

**Figure 4:** Electric conduction response of pristine graphene (dashed line) and graphene functionalised with a $V_2O_5$ layer (solid line) with respect to $NO_2$ and $NH_3$ gases. Measurements were done at room temperature under UV illumination. Horizontal bars indicate the time intervals of gas exposures. The inset depicts response to 100 ppm CO gas.

It should be noticed that after the PLD of $V_2O_5$ the conductivity of sensor decreases by a factor of 30. According to Raman spectra (Figure 1), the PLD process is fairly destructive and induces high concentration of defects to graphene 2D crystal lattice. These defects decrease the charge carrier mobility in graphene which is reflected in reduced electrical conductivity.

As compared to the pristine sensor, the responses to both gases are clearly improved after the functionalisation by PLD. The responses to 1 ppm $NO_2$ and 20 ppm $NH_3$ increase from 2 to 12%, and from 6.5% to 26%, respectively. The sensor current shows clear-cut and reversible responses to test gases $NO_2$ and $NH_3$. The conductivity of graphene sensors changes in opposite direction after inducing $NO_2$ or $NH_3$ into the test chamber. Graphene is typically a p-type conductor at ambient conditions, due to chemical doping by adsorbed oxygen and water molecules [27, 28]. Bearing in mind that $NH_3$ acts as a hole acceptor and $NO_2$ as a hole donor [1], the conductivity is expected to increase or decrease, respectively, that is indeed observed. Response to another common reducing gas, air pollutant carbon monoxide (CO) was also tested. The inset in Figure 4 demonstrates the response of $V_2O_5$ functionalised sensor to 100 ppm CO. Although the sensor is sensitive to CO and ~7 % decrease in conductivity is observed, the sensitivity to CO is much lower than to $NH_3$.

Figure 5 illustrates the effect of laser deposited $V_2O_5$ on $NH_3$ gas responses at different gas concentrations. The functionalisation with $V_2O_5$ clearly improves the performance of sensor. The response and recovery times of pristine and functionalised sensors for 8 ppm $NH_3$ gas were determined by fitting the time curves with the suitable functions. Either single or double exponential functions were used in the following form:

$$S(t) = \frac{\Delta I}{I_0} = A_0 + A_1 \exp\left[-(t-t_0)/t_1\right] \quad (1)$$

$$S(t) = \frac{\Delta I}{I_0} = A_0 + A_1 \exp\left[-(t-t_0)/t_1\right] + A_2 \exp\left[-(t-t_0)/t_2\right] \quad (2)$$

In equations (1) and (2) *S(t)* is relative change of conductance and $t_0$ is the initial moment of time, when the respective stepwise change in gas composition was introduced. Characteristic times $t_1$ and $t_2$ are related to the rate constants of adsorption and desorption of gas molecules onto pristine or $V_2O_5$ functionalised graphene. In principle, single exponential type of response (1) corresponds to the case where there is only one type of adsorption site available for adsorbing molecules at the surface of graphene sensor. Respectively, double exponential type of response (2) may describe the situation where two sites with different adsorption and desorption rate constants are available. The physical aspect of modelling of graphene sensor response is discussed more exhaustively in Jaaniso et al [29].

In case of pristine graphene, the response and recovery could be fitted with a single exponential function (1), which yielded characteristic times 578 s and 738 s for response and recovery, respectively. In case of $V_2O_5$ functionalised graphene, the response could be fitted with the double exponential function (2), with characteristic times 25 s and 175 s for response, and 42 s and 442 s for recovery. Although the response and recovery kinetics of functionalised sensor is double-exponential, it is safe to conclude that both the sensor response and recovery times become significantly faster as a result of laser deposition of $V_2O_5$ – the response was by ~3 times and the recovery by ~2 times faster for functionalised sensor. The amplitude of response during 600 s gas inlet increases by 2 times for 100 ppm, and by 8 times for 8 ppm $NH_3$, which suggests that the sensitivity is more significantly improved at lower concentrations.

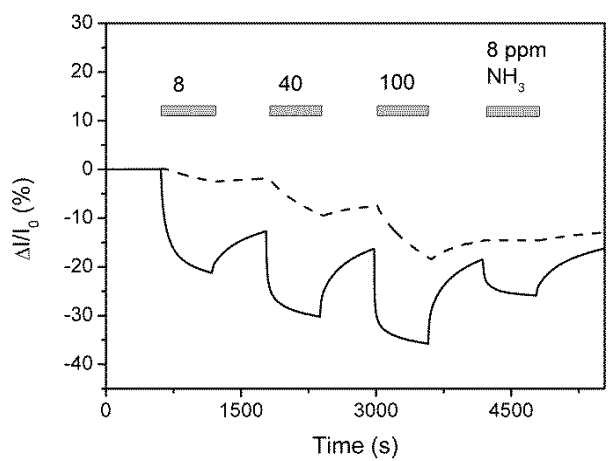

**Figure 5:** Response of the graphene sensor to different concentrations of $NH_3$ gas before (dashed line) and after (solid line) PLD functionalisation, measured under UV illumination at the room temperature. Bars show the duration of gas exposure.

# Discussion

Perfect graphene is relatively inert because of lack of dangling bonds and charged atoms on the surface. Consequently, the adsorption energy is due to van der Waals forces, and may be less or comparable to $k_B T$ ($k_B$ – Boltzmann constant, $T$ – absolute temperature) for gases at room temperature. Defects and dopant atoms in graphene can drastically increase both the adsorption of pollutant molecules, and the influence of gas adsorption on electronic properties of graphene [6, 30]. For instance, adsorption energy ($E_a$) of an $NH_3$ molecule on regular graphene is relatively small ($E_a$ ≤ 0.11 eV [6, 30]), but it is much higher for defective (up to 1.5 eV [6, 31]) or impurity-doped graphene (up to 1.4 eV [6, 30, 32]). Defects and doping atoms in graphene sheet are necessary to enhance the interaction between target gas molecules and graphene. However, this also results in increase of the interaction of $H_2O$ and $O_2$ molecules with graphene [30]. This would result in the situation where active desorption sites are already occupied by strongly bonding $O_2$ and $H_2O$ and the response to target gas would be slow and small. In some papers UV light or annealing in vacuum is used to clean the surface of graphene from strongly adsorbed gas molecules but during adsorption the excitation is not used [1, 26, 31]. However, in practical applications there is a constant need to activate the sensor surface by cleaning it from $H_2O$, $O_2$ and other adsorbates which would otherwise passivate it. The illumination by UV light as a tool for initiating or accelerating the desorption was introduced by Chen et al. [33] in case of carbon nanotube sensors but thereafter it was shown that the illumination is beneficial at the adsorption stage as well, considerably enhancing the magnitude of response [34]. Benefits of UV excitation are also clearly observed for single layer graphene functionalised by thin layer of laser deposited $ZrO_2$ or Ag. First, under UV the sensor responses were at least 2 to 3 times faster. Second, without UV illumination the recovery of signal in pure air was almost absent in case of $ZrO_2$, and only partial recovery of the signal occurred in case of Ag [14].

Generally, the graphene based sensors show much stronger response towards strongly oxidizing, paramagnetic (free radical) $NO_2$ molecule than that to any other gas, including $NH_3$. The binding energy is large and intermolecular charge transfer can occur when $NO_2$ adsorbs on either pristine, defective, or doped graphene [6, 31]. Table 1 compares $NO_2$ and $NH_3$ responses of our devices, and also of several other sensors reported in the literature, all based on a single layer graphene. Only qualitative discussion is possible, because in most cases the concentrations of gases are different. It can be seen that most sensors are more responsive to $NO_2$, as compared to $NH_3$. However, concerning the relative responses, the results of Yavari et al. [35] are comparable to our $V_2O_5$ functionalised sensor. The CVD grown graphene used in these experiments contain, according to the measured Raman spectra, relatively large concentration of defects of unspecified nature [35]. The role of graphene defects in case of $NH_3$ sensing is well illustrated by Lee et al. [31], where the response to $NH_3$ gas increased by 600% after creation of defects in the pristine defect-free graphene by reactive ion etching. At the same time, the response to $NO_2$ gas increased only by 33% [31]. We would like to point out that graphene sensors, functionalised by PLD with Ag and $ZrO_2$ in our previous work [14], showed a much larger response to 1 ppm $NO_2$ than to 20 ppm $NH_3$ (see Table 1, lines 2 and 3). Thus, decoration of graphene with $V_2O_5$ lends the sensor some degree of selectivity, and good sensitivity with respect to a reducing pollutant $NH_3$.

**Table 1:** Comparison of responses to $NO_2$ and $NH_3$ of the $V_2O_5$ functionalised graphene (Gr) gas sensor with other sensors based on pristine and defective graphene, or graphene functionalised with other materials.

| No. | Material | Response to $NO_2$ | Response to $NH_3$ | Reference |
|---|---|---|---|---|
| 1 | CVD Gr + $V_2O_5$ | 12 % (1 ppm) | 26 % (20 ppm) | This work |
| 2 | CVD Gr + $ZrO_2$ | 110 % (1 ppm) | 20 % (20 ppm) | This work[a] |
| 3 | CVD Gr + Ag | 20 % (1 ppm) | 11 % (20 ppm) | This work[a] |

| 4 | Exfoliated Gr | 4 % (1 ppm) | 4 % (1 ppm) | [1] |
| 5 | SiC/Gr + Au | 55 % (0.5 ppm) | 20 % (40 ppm) | [4] |
| 6 | Defective CVD Gr | 12 % (2 ppm) | 40 % (40 ppm) | [35] |
| 7 | Defective CVD Gr | 53 % (200 ppm) | 25 % (200 ppm) | [31] |
| 8 | B doped CVD Gr | 10 % (0.02 ppm) | 8 % (20 ppm) | [36] |

[a]Samples are described exhaustively earlier in [14]. Responses to $NH_3$ are measured in this work.

Introduction of dopants and clusters of atoms into graphene increases strongly the interaction with adsorbing molecules, due to increased charge transfer and formation of chemical bonds between the dopant and adsorbate [6, 32]. Moreover, changes in the electronic structure induced by adsorption of molecules are likely to modulate the conductivity of graphene [32].

According to Raman analysis shown in Figure 1, the PLD process induces a large number of defects in graphene. We propose that during functionalisation of graphene by PLD, the defect creation in graphene sheet by energetic plasma species is instantly followed by the $V_2O_5$ cluster growth at the defect site as the PLD process continues. The $V_2O_5$ clusters are probably chemically bound to defective sites of graphene sheet, yielding a strong and stable contact between the two phases. Chemical bonding is accompanied by charge transfer between $V_2O_5$ clusters and graphene. A similar situation was modelled previously by Lim and Wilcox [37] where large charge transfer and chemical bonding in platinum-graphene system were observed when Pt nanoclusters were chemically bonded to point defects of graphene sheet. Furthermore, adsorption of an $O_2$ molecule onto platinum nanocluster which was chemically bound to graphene resulted in large influence on the charge density distribution of the system [37].

The $V_2O_5$ is known by its catalytic properties, and as a good $NH_3$ adsorber [7, 8]. Good $NH_3$ gas sensing properties of $V_2O_5$ thin films and nanofibers have been demonstrated by Huotari et al. [12] and Modafferi et al. [38]. Two strongly bound adsorption species are typically observed as a result of reaction of $NH_3$ with two $V_2O_5$ adsorption sites, one with a surface OH group, forming positively charged $NH_4^+$, and the other with oxygen vacancy, forming species which is denoted as "coordinated $NH_3$". At the same time, the intensity of $V^{5+}=O$ related band in the infrared reflectance spectra decreases, which is an indication of the reduction of $V_2O_5$ [7, 9]. The corresponding ammonia oxidation reactions take place on $V_2O_5$ catalytic surface. A variety of redox processes are possible, for example:

$$2NH_3 + 3V_2O_5 \rightarrow N_2\uparrow + 3H_2O\uparrow + 3VO_2 \qquad (3)$$

Ambient oxygen can render the process reversible, as desired for sensor reset, e. g.:

$$2VO_2 + \tfrac{1}{2}O_2 \rightarrow V_2O_5 \qquad (4)$$

Obviously, strongly bound adsorbate species and the accompanying reduction of $V_2O_5$ can modify the charge distribution in the graphene-$V_2O_5$ system, thus modulating its conductivity.

According to Gao et al. [10], $NO_2$ can adsorb to $V_2O_5$ to form nitrato groups ($V^{5+}$–$NO_3$), and this process is reversible:

$$V^{5+}=O + NO_2 \rightleftharpoons V^{5+}\text{–}NO_3 \qquad (5)$$

As there are no redox reactions involved in this adsorption process the influence of $NO_2$ adsorption on conductivity of $V_2O_5$-graphene system is lower, as expected.

In addition to mechanisms described above, other energetically favourable adsorption sites can exist at phase boundaries or at the point defects in the parts of graphene that is exposed to the gas, since the $V_2O_5$ material is on average only 2.5 layers thick and can cover the graphene surface unevenly.

## Conclusion

The CVD graphene was functionalised by laser deposition of a sub-nanometer layer of catalytically active $V_2O_5$. The emergence of defect-related D and D' peaks and suppression of 2D peak in the Raman spectrum suggest that high concentration of defects was introduced during the deposition. According to XPS, the deposited vanadium was in the state of $V^{5+}$, which is the highest stable oxidation state of this element. As a result, an enhancement of sensing properties of graphene towards the reducing $NH_3$ gas was achieved. Improvement can arise from strong adsorption ability of $NH_3$ on $V_2O_5$ and possible redox reactions on the surface of $V_2O_5$, together with the activation of surface processes by using UV light illumination. The changing oxidation state of vanadium can modulate electrical conductivity of strongly coupled graphene-$V_2O_5$ system. Additionally, gas adsorption may be enhanced at the phase boundaries between a very thin nanostructured $V_2O_5$ layer and graphene substrate.

## Experimental

Graphene was grown on a commercial 25-µm thick polycrystalline copper foil (99.5%, Alfa Aesar) using a home-built CVD reactor. First the foil was annealed at 1000 °C in Ar/$H_2$ (both 99.999%, AGA Estonia) flow for 60 min, and then exposed to the mixture of 10% $CH_4$ (99.999%, AGA Estonia) in Ar at the same temperature for 120 min. The sample was let to cool slowly, 15°C/min in Ar flow. The as-grown graphene film was transferred onto Si/SiO$_2$ substrate by using poly(methyl methacrylate) (PMMA; MW ~997,000 Da, GPC, Alfa Aesar) as a supporting material. The PMMA solution (1 % in chlorobenzene) was spin coated on graphene/Cu, dried, and the Cu foil was dissolved in ammonium persulfate solution overnight. The floating PMMA/graphene film was rinsed with deionized water and transferred onto Si/SiO$_2$ substrate (see Figure 6) equipped with Pt electrodes (60 nm thick) that were deposited through a shadow mask by magnetron sputtering. The gap between the electrodes was 1 × 4 mm. The sample was dried in air and then heated on a hot plate to allow the PMMA film to soften that improved contact between graphene and the substrate. Then the PMMA layer was removed by dissolving in dichloromethane (Alfa Aesar).

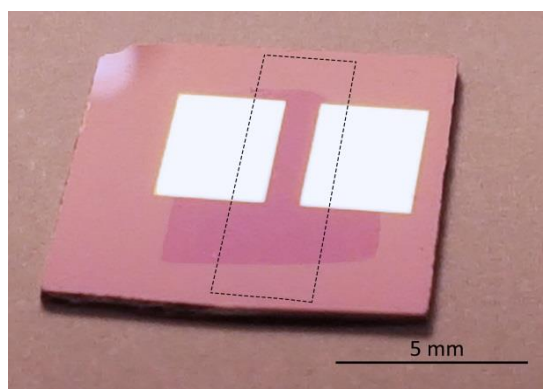

**Figure 6:** Photograph of a gas sensor device based on PLD-functionalised CVD graphene. The gap between the electrodes is 1 × 4 mm$^2$. The graphene sheet on top of the electrodes can be seen due to slightly different reflection properties. The area of laser deposited V$_2$O$_5$ is marked with dashed line.

In PLD process, a ceramic V$_2$O$_5$ pellet was used as an ablation target. The sensor substrate was held in place by a shadow mask through which V$_2$O$_5$ was deposited onto graphene. A KrF excimer laser (COMPexPro 205, Coherent; wavelength 248 nm, pulse width 25 ns) was used for deposition. For PLD target, fine microcrystalline powder of V$_2$O$_5$ (99.6 %, Aldrich) was pelleted at 740 MPa and subsequently baked at 550 °C in air for 5 hours. Before starting the PLD procedure, the PLD chamber was evacuated and the sensor substrates were heated in-situ at 150 °C for 1.5 hours and then cooled down to room temperature. The heating procedure was executed in order to clean the graphene surface and minimise the effect of traces of contaminants left from the manufacturing process of the graphene sensor. The $V_2O_5$ target was ablated by using laser pulse energy density of 5.0 J/cm$^2$ in the presence of $5 \times 10^{-2}$ mbar of $O_2$. During the deposition, the sensor substrates were kept at room temperature, with other typical process parameters being as follows: laser pulse repetition rate 5 Hz, number of laser pulses 120, and the distance between the substrate and the target 75 mm. Under exactly the same experimental conditions, another $V_2O_5$ layer was deposited onto a fused quartz substrate for the evaluation of mass thickness of the deposited layer in the XRF experiment.

Structural characterization of graphene was performed by using a micro-Raman spectroscopic system Renishaw inVia at the excitation wavelength of 514 nm. Vanadium concentration in the deposited layer was analysed with X-ray fluorescence device Rigaku ZSX 400. The oxidation state of deposited vanadium was also determined by X-ray fluorescence method. The XPS spectra were acquired using a SCIENTA SES-100 spectrometer. The excitation source was a non-monochromatic twin-anode x-ray tube (Thermo XR3E2) with Mg Kα (1253.6 eV) 300 W irradiation at the analyser-source angle of 45°. The semi-quantitative analysis was conducted using methods described by Seah et al. [39] for quantitative XPS measurements. The spectra are energy calibrated to 284.8 eV (from 284.9 eV) using the C-C peak component of C 1s. The software used for peak fitting was CasaXPS (version 2.3.16).

The measurements of electrical characteristics and gas sensitivity were carried out with sourcemeter (Keithley 2400) in a sample chamber with a volume of 7 cm$^3$ equipped with a gas mixing system based on mass flow controllers (Brooks, model SLA5820). The voltage applied between the electrodes was 100 mV. The gases used in our measurements $N_2$, $O_2$, $CO/N_2$, $NO_2/N_2$, and $NH_3/N_2$ were all 99.999% pure. Synthetic air of $N_2$ and $O_2$ mixture (79% and 21% in the mixture, respectively) was used as a carrier gas and the relative concentration of $N_2$ and $O_2$ gases was held constant during measurements. The gas flow through the sample chamber was kept constant at 200 sccm while the concentration of the test gas was regulated by the ratio of the flow rates of individual gas components. The nominal relative humidity of the testing gas was 20% during the measurements. All the sensor measurements were done at room temperature. The sensor system can be exposed to light from a Xe-Hg high pressure lamp (Hamamatsu). The UV wavelength of 356 nm was selected with a narrow-band interference filter (Andover). The light intensity on the sample was 10 – 20 mW/cm$^2$.

## Acknowledgements

The research leading to these results has received funding from the European Union's Horizon 2020 research and innovation programme under grant agreement No 649953, and from Estonian Research Council by institutional grants IUT34-27 and IUT2-24. The authors would like to thank Indrek Renge for valuable remarks and inspiring discussions.